\documentstyle[12pt]{article}
\newcommand{\la}{\langle}
\newcommand{\ra}{\rangle}
\begin{document}
\title{Analyses of multiplicity distributions 
by means of the Modified Negative Binomial Distribution 
and its KNO scaling function}
\author{Takeshi Osada, Noriaki Nakajima$^{1}$, Minoru Biyajima$^{2}$ \\and 
Naomichi Suzuki$^{3}$\\
${}^{}$Department of Physics, Tohoku University, Sendai 980-77, Japan \\
${}^{1} $RCNP, Osaka University, Ibaragi 567, Japan \\
${}^{2}$Department of Physics, Shinshu University, Matsumoto 390, Japan \\
${}^{3}$Matsusho Gakuen Junior College, Matsumoto 390-12, Japan }  
\date{\today}
\maketitle
\begin{abstract}
We analyze various data of multiplicity distributions by means of the 
Modified Negative Binomial Distribution (MNBD) and its KNO 
scaling function, since this MNBD explains the oscillating 
behavior of the cumulant moment observed in 
$e^+e^-$ annihilations, $h$-$h$ collisions and 
$e$-$p$ collisions. 
In the present analyses, 
we find that the MNBD (discrete distributions) describes the 
data of charged particles in $e^+e^-$ annihilations 
much better than the Negative Binomial Distribution (NBD). 

To investigate stochastic property of the MNBD, 
we derive the KNO scaling function from the discrete distribution 
by using a straightforward method and the Poisson transform. 
It is a new KNO function expressed by the Laguerre polynomials. 
In analyses of the data by using the KNO scaling function, 
we find that the MNBD describes the data better than the gamma function.

Thus, it can be said that 
the MNBD is one of useful formulas as well as NBD.
\end{abstract}

\section{Introduction}
Recently it has been found that cumulant moments of the multiplicity 
distributions in both $e^+e^-$ annihilations and hadronic collisions
show prominent oscillatory behavior when plotted as a function of their
order $q$ \cite{DREMIN}. In Ref.\cite{DREMIN} 
this behavior is attributed
to the QCD-type of branching processes apparently taking place in
those reactions. However, in Refs.\cite{NBS,NBSh} we have shown that
the same behavior of the moments emerges essentially from the modified 
negative binomial distribution (MNBD) (which actually describes
the data of {\it negatively charged particles} much better than the 
negative binomial distribution (NBD)
\cite{MNB1,MNB2,MNB3}). This distribution can be derived from the 
pure birth (PB) process with an initial condition given by the binomial 
distribution\cite{BSW}\footnote{
It is interesting to mention here that it comes also from the concept 
of purely bosonic sources as presented recently in Ref.\cite{BSWW}.}. \\

In this paper, first of all, we analyze 
various experimental data of 
{\it charged particles} \cite{HRS}-\cite{ua587} 
including $e$-$p$ collisions\cite{h1}
by the MNBD and the NBD, to elucidate why the MNBD describes the data 
better than the latter in $e^+e^-$ annihilations \cite{NBS,NBSh}.

Second, we derive the KNO scaling function of the MNBD
both by the straightforward method (i.e., proceeding to the limit 
of large multiplicities $n$ and large average multiplicities 
$\langle n\rangle$ while keeping the scaling variable
$z=n / \langle n\rangle $ finite and fixed) and by 
the Poisson transform.
Using this KNO scaling function we analyze the observed multiplicity
distributions in $e^{+}e^{-}$ annihilations, $h$-$h$ collisions and 
$e$-$p$ collisions. 

Finally the concluding remarks are given. 
The generalized MNBD is discussed in Appendix.

\section{Analyses of data by discrete distributions} 
It is known that the MNBD is obtained from the following equation 
governing the PB stochastic process\cite{MNB1,MNB2,MNB3,BSW}, 
\begin{eqnarray} 
dP(n,t)/dt = -\lambda n P(n,t) + \lambda(n-1)P(n-1,t) \label{PB-eq}
\end{eqnarray} 
with an initial condition
\begin{eqnarray} 
P(n,t=0)={}_{N}C_n ~\alpha^n(1-\alpha)^{N-n}, \qquad 
(0\le \alpha \le 1). 
\end{eqnarray}
Here $\lambda$ is a birth rate of particles. 
The parameter $t$ describes the evolution of the branching processes. 
The generating function (GF) of the distribution at $t=0$ is given by 
\begin{eqnarray} 
\Pi(u;t=0) \!\!&\equiv&\!\! \sum_{n=0}^{\infty}P(n,t=0)\times u^n\nonumber \\ 
\!\!&=&\!\! \sum_{n=0}^{\infty}
{}_NC_n \alpha^n(1-\alpha)^{N-n}u^n 
\nonumber \\
\!\!&=&\!\! [\alpha u +1 -\alpha]^N,
\end{eqnarray} 
The GF for the MNBD at 
$\displaystyle{t=\int_0^T\!\!dt=T}$ 
is given as Refs.\cite{NBS,NBSh,BSW},
\begin{eqnarray} 
\Pi(u;T)\!\!&\equiv&\!\! \sum_{n=0}^{\infty} P(n,T)\times u^n \nonumber \\
\!\!&=&\!\! \Big[ \frac{1-r_1(u-1)}{1-r_2(u-1)}\Big]^N,\label{GENE-MNBD}
\end{eqnarray} 
where 
\begin{eqnarray*} 
&&r_1=-\alpha(1+p)+p,\\
&&r_2=p~~(\ge 0),\\
&&p=\exp( \lambda T)-1, 
\end{eqnarray*} 
and $N$ (an integer) corresponds to the number of possible 
excited hadrons at the initial stage. 
The term $\lambda T$ is evaluated by 
$\displaystyle{\lambda T=\ln(1+r_2)}$.  
The MNBD is given by
the GF, Eq.(4), as 
\begin{eqnarray}
  P(0) &=& \Pi(0)= \left[ \frac{1+r_{1}}{1+r_{2}} \right]^N, \nonumber
            \\  
  P(n) &=& \frac{1}{n!}\frac{\partial^n \Pi(u)}{\partial u^n}= 
           \frac{1}{n!}\left(\frac{r_{1}}{r_{2}}\right)^N
           \left(\frac{r_2}{1+r_2}\right)^n
           \sum_{j=1}^{N} {}_{N}C_{j} \frac{\Gamma (n+j)}{\Gamma (j)}
           \left( \frac{r_2-r_1}{r_1} \right)^j
           \frac{1}{(1+r_2)^j},\nonumber \\ \label{Pn-MNBD}
\end{eqnarray}
\begin{eqnarray}
  r_{1} &=& \frac{1}{2} \left(C_{2}-1-\frac{1}{N} \right)\langle n \rangle-
           \frac{1}{2} ,
           \nonumber \\ 
  r_{2} &=& \frac{1}{2} \left(C_{2}-1+\frac{1}{N} \right)\langle n \rangle-
           \frac{1}{2}.  \label{eq:3} \nonumber 
\end{eqnarray}
Equation (\ref{Pn-MNBD}) is applied to various experimental 
data\cite{HRS}-\cite{h1}. 

For the sake of comparison with results obtained by Eq.(\ref{Pn-MNBD}), 
we also use the negative binomial distribution (NBD) given by
\begin{eqnarray} 
P(n)=\frac{\Gamma (n+k)}{\Gamma (n+1) 
\Gamma (k)} \left( \frac{k}{\langle n \rangle} \right)^{k} \left(1+ 
\frac{k}{\langle n \rangle} \right)^{-(n+k)}, \label{Pn-NBD}
\end{eqnarray} 
where $k>0$. 
The GF of the NBD is given by 
\begin{eqnarray}
\Pi(u)= \Big[~ 1-\frac{\langle n \rangle}{k}~(u-1)~\Big]^{-k}. 
\end{eqnarray} 
The GF, Eq.(4), of the MNBD reduces to that of the NBD, if $r_1=0$. 

Both the MNBD and the NBD are applied to analyses of the experimental
data in $e^+e^-$ annihilations\cite{HRS}-\cite{opal96}, 
$h$-$h$ collisions\cite{FNAL300}-\cite{ua587} and 
$e$-$p$ collisions\cite{h1}. 
Since there are maximum values of multiplicity observed, $n_{max}$, 
we introduce a possible bound for $N$ and 
truncate the multiplicity distribution $P(n)$ at $n_{max}$ and 
renormalize it 
as follows, 
\begin{eqnarray*} 
&&N\leq n_{max}~~,\\
&&C \times \sum_{n=0}^{n_{max}} P(n) = 1~. 
\end{eqnarray*} 
In analyses in Ref.\cite{MNB2,MNB3}, authors used the following 
treatments; 1) In the first case, 
the data $\langle n \rangle$ is used. 
$r_1$ and $N$ are free. 
Only a constraint, $r_2=r_1+\langle n \rangle/N$, 
is used. 2) In the second case, $\langle n \rangle$, $r_1$ and $N$ are 
free. $r_2=r_1+\langle n \rangle/N$ is used.

On the other hand we use $C_2$ and $\langle n \rangle$ as inputs. 
Only $N$ is free. 
Our results by means of Eqs.(\ref{Pn-MNBD}) and (\ref{Pn-NBD}) 
are given in Table I 
and some typical results of fitting are also given in Figs.1(a) and (b). 
It is found that in $e^+e^-$ annihilations, 
the minimum values of $\chi^2$'s  obtained by fitting 
Eq.(\ref{Pn-MNBD}) to the data are much smaller than those of the NBD.
This fact corresponds to our previous work in which 
we have found that 
the $H_q\equiv K_q/F_q$ moments obtained from the MNBD are much 
better than those of the NBD in $e^+e^-$ annihilation[2,3]: 
However in $h$-$h$ collisions they are almost equivalent. 

Because the data for $e$-$p$ collisions by H1 Collaborations\cite{h1} 
have recently reported, we calculate $H_q$ moments by the MNBD and the NBD. 
Two results of comparisons are shown in 
Figs.2(a) and (b) for energies $\sqrt{s}=$ 115-150 GeV and 
$\sqrt{s}=$ 150-185 GeV, respectively. 
It is found that the results of the MNBD and the NBD are almost 
equivalent as observed in $h$-$h$ collisions\cite{NBSh}.

\section{The KNO scaling function} 
\subsection{The straightforward method}

In order to know stochastic property of the MNBD, 
we consider the KNO scaling function of Eq.(\ref{Pn-MNBD}). 
Traditionally the KNO scaling function is derived from the multiplicity
distribution $P(n)$ multiplied by the corresponding mean multiplicity 
$\la n\ra$ by going to the large multiplicity $n$ and large mean 
multiplicity $\la n\ra$ limit while keeping their ratio, 
$ z = \lim_{n,\langle n \rangle\rightarrow 
\infty}n/\langle n \rangle$ fixed. In our case, starting 
from Eq.(\ref{Pn-MNBD}) we arrive at the following function 
\begin{eqnarray}
        \Psi(z)  & \equiv & \lim_{n,\langle n \rangle\rightarrow 
\infty}\langle n \rangle P(n)
                       \nonumber \\
                 & = & \left(\frac{r'_1}{r'_2}\right)^{N}
             e^{-\frac{\langle n \rangle}{r'_2}z}
             \sum_{j=1}^{N}{}_NC_{j} \frac{1}{\Gamma(j)}
             \left(\frac{r'_2-r'_1}{r'_1}\right)^{j}
              \left(\frac{\langle n \rangle}{r'_2} \right)^{j}z^{j-1}.
        \label{eq:5}
\end{eqnarray}
The parameters $r'_1$ and $r'_2$ in Eq.(\ref{eq:5}) are given by
\begin{eqnarray}
  r'_{1} &=& 
    \frac{1}{2} \left(C_{2}-1-\frac{1}{N} \right)\langle n \rangle, 
           \nonumber \\ 
  r'_{2} &=& 
    \frac{1}{2} \left(C_{2}-1+\frac{1}{N} \right)\langle n \rangle,
 \label{eq:06}
\end{eqnarray}
which are slightly different from 
$r_1$, $r_2$ in Eq.(\ref{Pn-MNBD}), because $\la n \ra \gg 1$.
It should be noticed that the normalization of Eq.(\ref{eq:5}) 
differs from the unity,
\begin{equation}
        \int_{0}^{\infty}\Psi(z)dz = 1 -
               \left(\frac{r'_1}{r'_2}\right)^{N},
        \label{eq:07}
\end{equation}
where the second term corresponds to the term 
$\langle n \rangle P(0)$ in Eq.(\ref{Pn-MNBD}).\\
\subsection{The Poisson transform}

The KNO scaling function $\Psi(z,t)$ is related to the 
multiplicity distribution function $P(n,t)$ 
by the Poisson transform\cite{SB}\\

\begin{picture}(400,30)(0,0)
       \put(280,10){$\Psi(z,t)$ .}
       \put(120,10){$P(n,t)$}
       \put(160,14){\vector(1,0){110}}
       \put(270,10){\vector(-1,0){110}}
       \put(180,18){\scriptsize  inverse  Poisson   trans.}
       \put(190,2){\scriptsize  Poisson   trans.}
 \end{picture}

\noindent
In this approach we obtain that 
\begin{eqnarray} 
&&     P(n,t)=\int_{0}^{\infty}
               \frac{(\alpha \omega)^{n}}{n!}e^{-\alpha \omega}
               \Psi \left( \frac{\omega}{ \la n \ra /\alpha},t \right)
               \frac{d \omega}{ \la n \ra / \alpha} ,\\ 
&&    \Psi\left( \frac{ \omega}{ \la n \ra / \alpha},t \right)
           = \frac{1}{2 \pi}e^{\alpha \omega}
               \frac{\la n \ra }{\alpha}
               \int_{-\infty}^{\infty}
               e^{-ix \omega}
               \sum_{n=0}^{\infty} \left( \frac{ix}{\alpha} 
               \right)^n P(n,t)dx            
         \nonumber \\  
&&\hspace*{2.5cm}  = \frac{1}{2\pi i}\int_{\sigma-i\infty}^{\sigma+i\infty}
             e^{sz}
             \Pi(1-\frac{s}{\langle n \rangle},t)ds,
\end{eqnarray}
where $\Pi(u,t)$ is the generating function of $P(n,t)$. 
These equations hold also at $t=0$ (the stationary function).\\ 

Using the generating function Eq.(\ref{GENE-MNBD}), 
the KNO scaling function $\Psi(z)$ is given by 
the following inverse Laplace transform
\begin{eqnarray}
   \Psi \left( z\right)
         = \frac{1}{2\pi i}\int_{\sigma-i\infty}^{\sigma+i\infty}
             e^{sz}
             \left(\frac{r_1}{r_2}\right)^{N}
             \sum_{j=0}^{N}{}_{N}C_{j}
           \left( \frac{r_2-r_1}{r_1}\frac{\langle n \rangle}{r_2} \right)^{j}
             \left( s+\frac{\langle n \rangle}{r_2}\right)^{-j}ds.
                \label{eq:10}
\end{eqnarray}
Then we arrive at the KNO scaling function for the MNBD,
\begin{eqnarray}
  \Psi(z) &=& \left(\frac{r'_1}{r'_2}\right)^{N}\delta(z-\epsilon) +
             \left(\frac{r'_1}{r'_2}\right)^{N}
             e^{-\frac{\langle n \rangle}{r'_2}z}
             \sum_{j=1}^{N}{}_NC_{j} \frac{1}{\Gamma(j)}
             \left(\frac{r'_2-r'_1}{r'_1}\right)^{j}
              \left(\frac{\langle n \rangle}{r'_2} \right)^{j}z^{j-1}
          \nonumber \\
          &=& \left(\frac{r'_1}{r'_2}\right)^{N}\delta(z-\epsilon) +
              \left(\frac{r'_1}{r'_2}\right)^{N}
              \frac{r'_2-r'_1}{r'_2}\frac{\langle 
              n \rangle}{r'_1}
              e^{-\frac{\langle n \rangle}{r'_2}z}
              L^{(1)}_{N-1}
              \left(-\frac{r'_2-r'_1}{r'_2}\frac{\langle 
              n \rangle}{r'_1}z \right),
                \label{MNBD-KNO}
\end{eqnarray}
where 
$L_{n}^{(\alpha)}(x)$ is the associated Laguerre's polynomial.
In Eq.(\ref{MNBD-KNO}) the first term corresponds to the constant term 
in Eq.(\ref{eq:10})~(, or $\langle n \rangle P(0)$).
In the numerical calculations the first term is very small 
because $(r_1'/r_2')^N$ where $(N >1)$. 
Therefore Eqs.(7) and (13) are almost equivalent in numerical analyses. 
The generalized MNBD is discussed in Appendix. 

\clearpage
\noindent 
The $C_q$ moment of the MNBD is given by
\begin{eqnarray} 
C_q \!\!&\equiv&\!\! \int_0^\infty dz z^q \Psi(z) \nonumber \\ 
 \!\!&=&\!\!\left(\frac{r'_1}{r'_2}\right)^{N}
             e^{-\frac{\langle n \rangle}{r'_2}z}
             \sum_{j=1}^{N}{}_NC_{j} \frac{1}{\Gamma(j)}
             \left(\frac{r'_2-r'_1}{r'_1}\right)^{j}
              \left(\frac{\langle n \rangle}{r'_2} \right)^{j}z^{j-1+q}
          \nonumber \\
\!\!&=&\!\! 
\left(\frac{r'_1}{r'_2}\right)^{N} 
\left(\frac{r'_2}{\langle n \rangle}\right)^{q}
\sum_{j=1}^{N}{}_NC_{j} \frac{\Gamma(j+q)}{\Gamma(j)}
             \left(\frac{r'_2-r'_1}{r'_1}\right)^{j}
\end{eqnarray}

We also analyze the data by the gamma function 
which is the KNO scaling function obtained from Eq.(\ref{Pn-NBD}); 
\begin{equation}
\Psi^{NBD}(z)=\frac{k^k}{\Gamma (k)}e^{-kz}z^{k-1}.\label{KNO-NBD}
\end{equation}

\subsection{Analysis of the experimental data}
We investigate the applicability of the MNBD 
as presented in its KNO form (i.e., using Eq.(\ref{MNBD-KNO}))
to the description of the observed multiplicity distributions 
in $e^{+}e^{-}$ annihilations\cite{HRS}-\cite{opal96}, 
$h$-$h$ collisions\cite{FNAL300,ua587} 
and $e$-$p$ collisions\cite{h1}. 
In actual analysis the values of the $r_1'$ and $r_2'$ are 
determined by using the experimental data of $C_2$ and 
$\langle n \rangle$ as menstioned before. 
Table II shows obtained parameters $N$ in Eq.(\ref{MNBD-KNO}) 
and minimum values of $\chi^2$'s of fitting
to the experimental data. 
See Figs.3(a) and (b). We show here some typical results of fittings. 
The results using Eq.(\ref{KNO-NBD}) are also 
shown there. 
As is seen from Table II, the minimum values of $\chi^{2}$'s 
for the MNBD fitting 
are a little smaller than those for the NBD fitting in low energies 
$e^{+}e^{-}$ annihilations and $h$-$h$ collisions 
(below $\sqrt{s}\sim $ 50~GeV). 
For $e$-$p$ collisions the minimum values of $\chi^2$'s are 
also smaller than those of the NBD up to energy $\sqrt{s}$=220~GeV.

On the other hand, in high-energy 
(LEP and SPS energies for the $e^{+}e^{-}$ annihilations and $h$-$h$ 
collisions, respectively), 
values of 
the minimum $\chi^2$'s by the MNBD fitting are almost equivalent 
to those of the NBD fitting.\\

\section{Concluding remarks} 

Through analyses of the data by discrete distributions, we find that 
the MNBD describes the data in $e^+e^-$ annihilation much better than 
the NBD. See Table I. 
We find that $N=7\sim 13$ except for the data by 
HRS and TASSO Collabs. 
On the other hand, in \cite{MNB2,MNB3} $N= 7\sim 8$ for 
\underline{{\it negatively}} charged particles is obtained, 
provided that their treatments for parameters are used.
In other words, we have to pay our attention to the 
determination of parameters contained in the formulas.

Second the $H_q$ moments of $e$-$p$ collisions by H1 Collab. are 
analyzed by two distributions. The results are almost the same as 
$h$-$h$ collisions. 
The $\chi^2$~'s in Table I are slightly better 
than those in Refs.\cite{MNB2,MNB3}, 
when our treatment for parameters is used.

Third, the KNO scaling function of the MNBD is obtained. 
As far as our knowledge for the KNO scaling functions\cite{SALEH} 
is correct, we have not known the KNO scaling function expressed by 
the Laguerre polynomials. It is applied to the analyses of the 
observed multiplicity distribution 
in $e^{+}e^{-}$ annihilations, $h$-$h$ collisions, and $e$-$p$ collisions.
The data are also analyzed by the gamma distribution. 
As is seen from Table II, 
the MNBD describes the experimental data 
better (for $e^+e^-$ annihilations) than 
or as well as (for $h$-$h$ and $e$-$p$ collisions ) the NBD. 

In conclusion, it can be stressed that the MNBD is useful 
as well as the NBD. See also Ref.{\cite{SZWED}}. 

What we have found in the present analyses suggests us the following: 
In $e^+e^-$ annihilation, the stochastic pure birth process 
with the binomial distribution at $t$=0 is useful, 
becasuse the finite number (corresponding to $N=7 \sim 13$) 
of the excited hadrons (or the pair creation of quarks) 
is probably expected. 
In other words, the binomial distribution 
as the initial condition is more realistic than other 
condition~($\delta_{n,k}$) in the stochastic approach.

To know stochastic property of the MNBD, we have 
considered the generalized MNBD.
In its concrete application, we have known 
that the discrete distribution 
of Eq.(A5) can not explain
the oscillating behaviors of the cumulant moment observed in 
$e^+e^-$ annihilations and in hadronic collisions 
\underline{{\it much better}} 
 than the MNBD, Eq.(\ref{Pn-MNBD}).
For example, $\chi^2$'s values for discrete analysis of data 
at $\sqrt{s}$=900 GeV are as follows:
($N,k,\chi^2$)=(1, 2.22, 65.2), (2, 1.33, 63.0) and 
(3, 0.06, 60.8). These results suggest us that it is difficult to 
determine the best combination of parameters, because of 
three parameters. It seems to be necessary more carefulness 
and skillfulness than the use of the MNBD. 
Thus the generalized MNBD is given in Appendix. 
However, we are expecting that Eq.(A5) 
will become useful in analyses of data of some reactions at higher energies, 
since it has both stochastic characteristics of the MNBD and the NBD.

\vspace{5mm}

\noindent 
{\bf Acknowledgments:~~~}
One of the authors (M. B.) is partially supported by the Grant-in Aid 
for Scientific Research from the Ministry of Education, Science, 
Sports and Culture (No. 06640383) and (No. 09440103).
 T.O. would like to thank those who supported him at the Department of 
Physics of Shinshu University. 
N. S. thanks for the financial support by Matsusho Gakuen Junior College. 
We are grateful to G. Wilk for his reading the manuscript.

\vspace*{1cm}

\noindent 
{\bf Appendix:~~~} \\
In order to know the stochastic structure of the MNBD in detail, 
we discuss the following point: The solution obtained from the 
branching equation of the pure birth process with the immigration 
under the initial condition of the binomial distribution\cite{NBSh,BSW} 
is one of the extensions of both the MNBD and the NBD.~ 
In this case the stochastic equation (\ref{PB-eq}) should be 
changed as follows: 
$$
dP(n,t)/dt = -\lambda n P(n,t) + \lambda(n-1)P(n-1,t) 
             -\lambda_0 P(n,t)+\lambda_0 P(n-1,t), \eqno{(A1)}
$$
where $\lambda_0$ is an immigration rate. 
Its generating function is given as
$$
	\Pi(u) = \sum_{n=0}^{}P(n)u^{n}
	       = [1-\tilde{r}_1(u-1)]^{N}[1-\tilde{r}_2(u-1)]^{-k-N}.
	\eqno{(A2)}
$$
where 
$$
   \tilde{r}_{1}=\frac{\langle n \rangle}{k}\left \{ 1 -\sqrt{\frac{k+N}{N}
            \left[ -k \left( C_2 -1-\frac{1}{\langle n \rangle} \right) 
            +1 \right] }~~\right \}
\eqno{(A3)}
$$
$$
    \tilde{r}_{2}= \frac{N \tilde{r}_1 + \langle n \rangle}{k+N},
\eqno{(A4)}
$$
and $k=\lambda_0/\lambda$. 
The MNBD is obtained by neglecting the power 
$k$ in $\Pi(u)$. The generating function of the NBD 
is given by neglecting the power $N$ in Eq.(A2).
The physical meaning of the immigration term
$k$ may be interpreted as a possible contribution from 
constituent quarks and gluons. 
Using Eq.(\ref{GENE-MNBD}), we have directly the KNO scaling function for 
Eq.(A2)
\\
$$  \Psi(z) =   \left(\frac{\tilde{r}_1}{\tilde{r}_2}\right)^{N}
                 e^{-\frac{\langle n \rangle}{\tilde{r}_2}z}
                 \sum_{j=0}^{N}{}_NC_{j} \frac{1}{\Gamma(k+j)}
                 \left(\frac{\tilde{r}_2-\tilde{r}_1}{\tilde{r}_1}\right)^{j}
                 \left(\frac{\langle n \rangle}{\tilde{r}_2} 
                 \right)^{k+j}z^{k+j-1}
$$
$$\hspace*{2.5cm}      =  \frac{\Gamma(N+1)}{\Gamma(N+k)}
                 \left(\frac{\tilde{r}_1}{\tilde{r}_2}\right)^{N}
                 \left(\frac{\langle n \rangle}{\tilde{r}_2}\right)^{k}
                 z^{k-1}
                 e^{-\frac{\langle n \rangle}{\tilde{r}_2}z}
                 L^{(k-1)}_{N} \left( -\frac{\tilde{r}_2-
                 \tilde{r}_1}{\tilde{r}_2}\frac{\langle 
                 n \rangle}{\tilde{r}_1}z\right).
                 \eqno{(A5)}\label{eq:12}
$$
Here the term $1/\langle n \rangle$ in the parenthesis in 
the squared root of Eq.(A3) is neglected because 
$\langle n \rangle \to \infty$. 
It also noticed that all factors $\langle n \rangle$'s in Eq.(A5) 
are canceled  out. 
This function becomes the KNO scaling function of the MNBD 
Eq.(\ref{MNBD-KNO}) when $k \rightarrow 0$, and reduces to the gamma 
distribution Eq.(\ref{KNO-NBD}) if $N=0$.


%
\clearpage 
\begin{flushleft}
{\large Table captions}
\end{flushleft}

\noindent 
{\bf Table I}\\ 
The minimum values of $\chi^2$'s for fitting to the experimental 
data using the MNBD and the NBD, respectively.
The data of $\langle n \rangle$ and $C_2$ are used. 
We also show $N$ giving the minimum values of $\chi^2$'s in Eq.(6). 
In analysis of the data by UA5 Collab ($\sqrt{s}$=900 GeV ), 
we find a minimum $\chi^2$ at $N$=3 
giving $P(n)<0$ at small integer n. 
We discard this and adopt $N=4$ giving $P(n)>0$ for all $n$. 
In analysis of the data by DELPHI Collab. 
$\langle n \rangle =21.26$  and $C_2=1.0908$ are calculated 
from the data. \\
The symbole $\ast$ denotes the minimum value of $\chi^2$'s 
between 2 columns. $\chi^2$'s in Ref.\cite{MNB3} are 
obtained by means of the statistical errors only.\\
\\

\noindent 
{\bf Table II}\\ 
The same as Table I but for fitting of the KNO scaling function 
obtained from the MNBD and the NBD, respectively. 
The data of $\langle n \rangle$ and $C_2$ are used. 
In analysis of the data by OPAL ($\sqrt{s}$=133 GeV ) and 
UA5 ($\sqrt{s}$ =900 GeV) Collabs, we find minimum $\chi^2$'s at 
$N=8$ and $N=3$ giving $P(n)<0$ in small $z$, respectively. 
We discard these and adopt $N=10$ and $N=4$ 
giving $P(n)>0$ for all $n$ for the data by OPAL and UA5 Collabs, 
respectively.\\
The symbol $\ast$ denotes the minimum value of $\chi^2$'s 
between 2 columns.

%
%
\vspace*{1cm}
\begin{flushleft}
{\large Figure captions}
\end{flushleft}
{\bf [Fig. 1]~} Results obtained by discrete distributions. \\
\noindent 
(a)~The result of fitting by Eqs.(5) and (7) to the data observed 
    by TASSO Collab \cite{tass89}.\\~~~($\sqrt{s}$=34.8 GeV.)\\
(b)~The same as (a) but for the data observed by OPAL Collab \cite{opal96}.
    ($\sqrt{s}$=91.2 GeV.)\\

\noindent 
{\bf [Fig.2(a) and (b)]~} Analyses of $H_q$ moment in $e$-$p$ collisions 
by H1 Collaboration for energies (a)~$\sqrt{s}$=115-150 GeV
and (b)~$\sqrt{s}$=150-185 GeV.\\

\noindent 
{\bf [Fig. 3(a) and (b)]~}
 Results obtained by the KNO scaling functions.\\
The same as Figs.1(a) and (b) but for fittings Eqs.(\ref{MNBD-KNO}) 
and (\ref{KNO-NBD}). 
%

\newcommand{\lw}[1]{\smash{\lower2.0ex\hbox{#1}}}
\begin{center}
\begin{tabular}{|ll|r|lr||r|}
\hline 
Exp. &[GeV]&$n_{max}$&\multicolumn{2}{c||}{MNBD} &\multicolumn{1}{c|}{NBD}\\
 \hline \hline 
TASSO &14  &26&[N=9~]&*9.34/10 &25.0/11\\ \hline  
TASSO &22  &28&[N=11]&*2.14/11 &10.1/12\\  \hline 
HRS   &29  &28&[N=20]&*6.24/11 &8.20/12\\  \hline
TASSO &34.8&36&[N=13]&*11.8/15 &39.8/16\\  \hline 
TASSO &43.6&38&[N=38]&*7.42/16 &*7.40/17\\ \hline  
AMY   &50. &38&[N=10]&*2.75/16 &14.6/17\\ \hline
AMY   &57. &40&[N=12]&*9.84/17 &51.8/18\\ \hline
AMY   &60. &40&[N=12]&*5.79/17 &11.4/18\\ \hline
SLD   &91.2&50&[N=13]&*27.1/21 &154./22\\ \hline
L3    &91.2&50&[N=9~]&*15.0/20 &30.6/21\\ \hline
ALEPH '91 &91.2&48&[N=11]& *9.07/22&12.4/23 \\ \hline 
ALEPH '95 &91.2&54&[N= 9]& *3.42/25&8.97/25 \\ \hline 
DELPHI&91.2&52&[N=10]&*14.2/22 &185./23\\ \hline
OPAL  &91.2&54&[N=9~]&*4.03/24 &47.5/25\\ \hline
OPAL  &133.&54&[N=7~]&*4.57/21 &39.6/22\\ 
\hline \hline 
FNAL  &23.9&28&[N=28]&*27.1/11&34.1/12\\ \hline  
ISR   &30.4&34&[N=5~]&*24.6/14 & 36.8/15\\ \hline 
FNAL  &38.8&32&[N=32]&*55.7/13&81.0/14\\ \hline   
ISR   &44.5&38&[N=5~]&*5.27/16 & 16.6/17\\ \hline
ISR   &52.6&42&[N=7~]&*6.41/18 &*6.96/19\\ \hline
ISR   &62.2&40&[N=40]&*24.9/17 & 29.2/18\\ \hline
UA5   &200 &64&[N=4~]&*7.72/28 & 9.87/29\\ \hline
UA5   &546&112&[N=4~]& 79.1/44 &*76.1/45\\ \hline
UA5   &900&122&[N=4~]& 84.6/51 &*77.5/52\\ 
\hline \hline 
H1             & 80-115&\lw{18}&[N=18]& *2.93/16 & 2.92/17  \\ \cline{4-6}
$1<\eta^{*}<5$ & Ref.[6]& &[N=7] &  2.70/18   &        \\ \cline{2-6}
               &115-150&\lw{21}&[N=21]& 8.52/19  & *7.01/20 \\ \cline{4-6}
               & Ref.[6]& &[N=7] &  10.6/21   &        \\ \cline{2-6}
               &150-185&\lw{22}&[N=22]& *5.12/20 & 5.65/21  \\ \cline{4-6}
               & Ref.[6]& &[N=7] &  4.60/22   &        \\ \cline{2-6}
               &185-220&\lw{23}&[N=23]& *5.99/21 & 8.97/22  \\ \cline{4-6}
               & Ref.[6]& &[N=7] &  11.3/23   &        \\
\hline  
\end{tabular}
\end{center} 
\begin{center}
{\bf \large Table I} 
\end{center} 

\begin{center}
\begin{tabular}{|l l|l r||r|}
	\hline
	Exp. & GeV & \multicolumn{2}{c||}{MNBD} &\multicolumn{1}{c|}{NBD}\\
	\hline
	\hline
	TASSO & 14 &[N=29]      & *10.9/11 &19.0/12 \\
	\hline
	TASSO & 22 &[N=28]& *3.53/12 & 15.3/13 \\
	\hline
	HRS   & 29 &[N=28]& *15.0/12 &  37.4/13 \\
	\hline
	TASSO &34.8&[N=35]     & *8.85/16 &  42.7/17 \\
	\hline
	TASSO &43.6&[N=38]& *14.1/17 &  49.9/18 \\
	\hline
	AMY   & 50 &[N=38]& *4.19/17 &  *4.45/18 \\
	\hline
	AMY   & 57 &[N=29]     & *7.82/18 &  19.1/19 \\
	\hline
	AMY   & 60 &[N=19]     & *5.35/18 &  8.71/19 \\
	\hline
	SLD   &91.2&[N=19]     & *35.3/22 &  200./23\\
	\hline
	L3    &91.2&[N=11]     & *13.3/21 &  *13.4/22 \\
	\hline
	ALEPH '91 &91.2&[N=15] & *8.51/22 & 9.95/23 \\
	\hline
	ALEPH '95 &91.2&[N=10] & *3.29/24 & *3.59/25 \\
	\hline
	DELPHI&91.2&[N=12]     & *17.1/23 &  21.0/24 \\
	\hline
	OPAL  &91.2&[N=11]     &  *4.46/25 &  *4.39/26 \\
	\hline
	OPAL  &133 &[N=10]     &  11.2/22 &  *9.63/23 \\
	\hline
	\hline
	FNAL  &23.9&[N=28]& *26.7/12 &  77.1/13 \\
	\hline
	ISR   &30.4&[N=6~]      & *21.0/15 &  23.1/16\\
	\hline
	FNAL  &38.8&[N=32]& *45.7/14 &  131./15\\
	\hline
	ISR   &44.5&[N=6~]      & *5.38/17 & 8.19/18\\
	\hline
	ISR   &52.6&[N=8~]      & *5.94/19 &  33.9/20\\
	\hline
	ISR   &62.2&[N=40]& *23.5/18 &  59.7/19 \\
	\hline
	UA5   &200 &[N=4~]      &  *7.70/29 &  *7.69/30\\
	\hline
	UA5   &546 &[N=4~]      & 67.6/45 &  *59.5/46\\
	\hline
	UA5   &900 &[N=4~]      & 78.2/52 &  *63.5/53\\
	\hline
	\hline
H1     &80-115 &[N=18]& *20.5/17 & 37.7/18 \\
    \cline{2-5}
 $1<\eta^{*}<5$&115-150&[N=21]& *19.5/20 &32.8/21 \\
	\cline{2-5}
&150-185 & [N=22] & *9.55/21 &18.4/22  \\
	\cline{2-5}
&185-220 & [N=23] & *12.3/22 & 23.3/23 \\
	\hline
\end{tabular}
\end{center}
\begin{center}
{\bf \large Table II} 
\end{center} 


\begin{thebibliography}{cc}
 \bibitem{DREMIN}I. M. Dremin and V. A. Nechitailo, JETP Lett. 
                 {\bf 58} (1993) 881; 
                 I. M. Dremin and R. Hwa, Phys. Rev.
                 {\bf D49} (1994) 5805; 
                 I. M. Dremin, Phys. Lett. {\bf B341}
                 (1994) 95.
 \bibitem{NBS} N. Suzuki, M. Biyajima and N. Nakajima, Phys. Rev.
               {\bf D53} (1996) 3582 and {\bf D54} (1996) 3653. 
 \bibitem{NBSh} N. Nakajima, M. Biyajima and N. Suzuki, Phys. Rev. 
               {\bf D54} (1996) 4333. 
 \bibitem{MNB1} P. V. Chliapnikov and O. G. Tchikilev, Phys. Lett.
               {\bf B242} (1990) 275.
 \bibitem{MNB2} P. V. Chliapnikov, O. G. Tchikilev, and V. A. Uvarov, 
                Phys. Lett. {\bf B352} (1995) 461. 
 \bibitem{MNB3} O. G. Tchikilev,  Phys. Lett. {\bf B382} (1996) 296,
~{\it ibid} {\bf 388}(1996),848.
 \bibitem{BSW} N. Suzuki, M. Biyajima and G. Wilk, Phys. Lett. 
               {\bf B268} (1991) 447. 
 \bibitem{BSWW} M. Biyajima, N. Suzuki, G. Wilk and Z. W\l odarczyk.
                 Phys. Lett. {\bf B386} (1996) 279.
 \bibitem{HRS}HRS Collab., M. Derrick et al., Phys. Rev. 
                 {\bf D34} (1986) 3304.
 \bibitem{tass89}TASSO Collab., W. Braunschweig et al., Z. Phys. 
                 {\bf C45} (1989) 193.
 \bibitem{amy}AMY Collab., H. W. Zheng et al., Phys. Rev. 
                 {\bf D42} (1990) 737.
 \bibitem{aleph91} ALEPH Collab., D. Decamp et al., Phys. Lett. 
                 {\bf B273} (1991) 181. 	
 \bibitem{aleph95} ALEPH Collab., D. Buskulic et al., 
                   Z. Phys. {\bf C69} (1995) 15. 	
 \bibitem{delp91}DELPHI Collab., P. Abreu et al., Z. Phys. 
  {\bf C52} (1991) 271.
 \bibitem{opal92}OPAL Collab., P. D. Acton et al., Z. Phys.  
 {\bf C53} (1992) 539.
 \bibitem{opal96}OPAL Collab., P. D. Acton et al., CERN-PPE/96-47.
 \bibitem{FNAL300}A. Firestone et al., Phys. Rev. {\bf D}10(1974) 2080. 
 \bibitem{FNAL800}E743 Collab. R. Ammer et al., 
  Phys .Lett.{\bf B}178(1986) 124. 
 \bibitem{isr}ISR Collab., A. Breakstone et al., Phys. Rev. 
                 {\bf D30} (1984) 528.
 \bibitem{ua587} UA5 Collab., G. J. Alner et al., Phys. Rep. 
{\bf 154} (1987) 247; UA5 Collab., R. E. Ansorge et al., Z. Phys.  
 {\bf C43} (1989) 357. 
 \bibitem{h1} H1 Collab., S. Aid et al., DESY preprint 96-160.
 \bibitem{SB} M. Biyajima, Prog. Theor. Phys. {\bf 69}(1983) 966;
            {\it ibid} {\bf 70}(1983) 1468A.
 \bibitem{Giovannini} A. Giovannini, S. Lupia and R. Ugoccini, 
 Phys. Lett. {\bf B374}, (1996) 231. 
 F. Becattini, A. Giovannini and S. Lupia, Z. Phys. {\bf C72} (1996), 491. 
 \bibitem{SALEH} S. Saleh, `Photoelectron Statistics',  
(Springer-Verlag, Berlin, 1998). 
 \bibitem{SZWED} R. Szwed, G. Wrochna, and A.K. Wr\'oblewski, 
Mod. Phys. Lett. {\bf A6}, (1991), 245. 
\end{thebibliography}
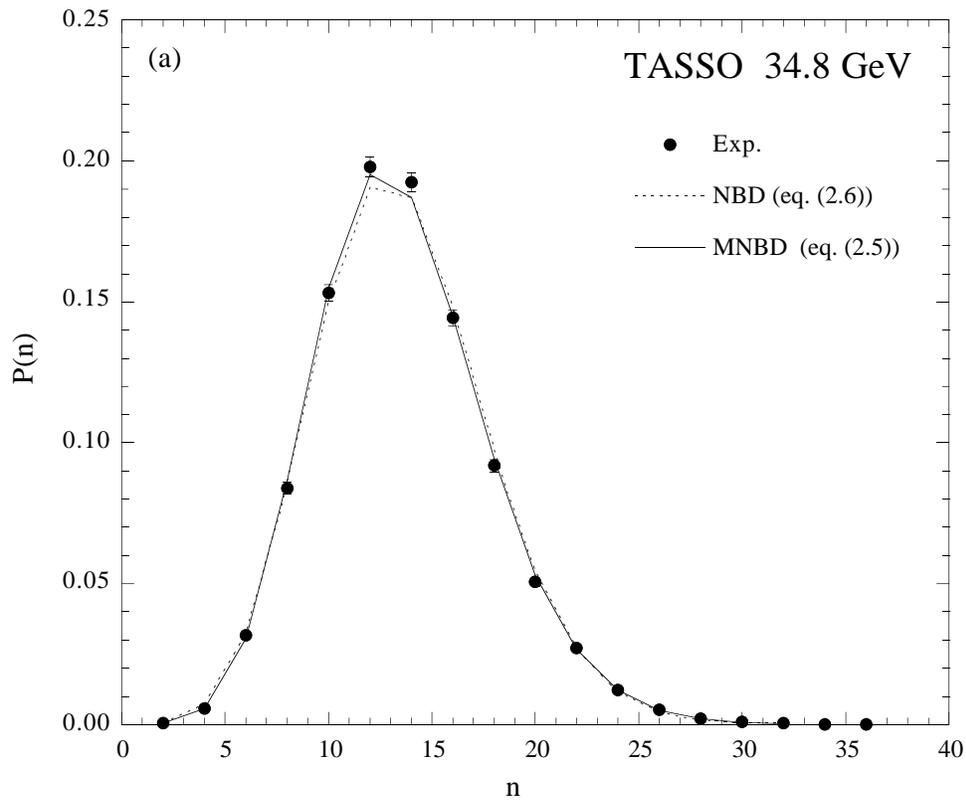
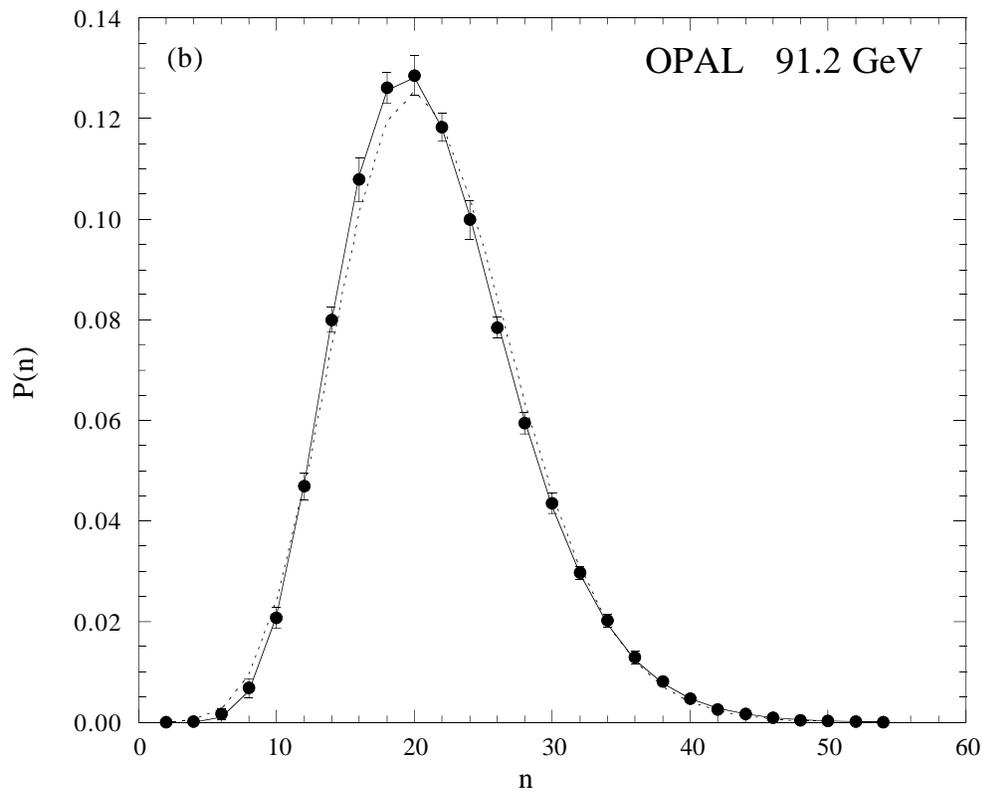
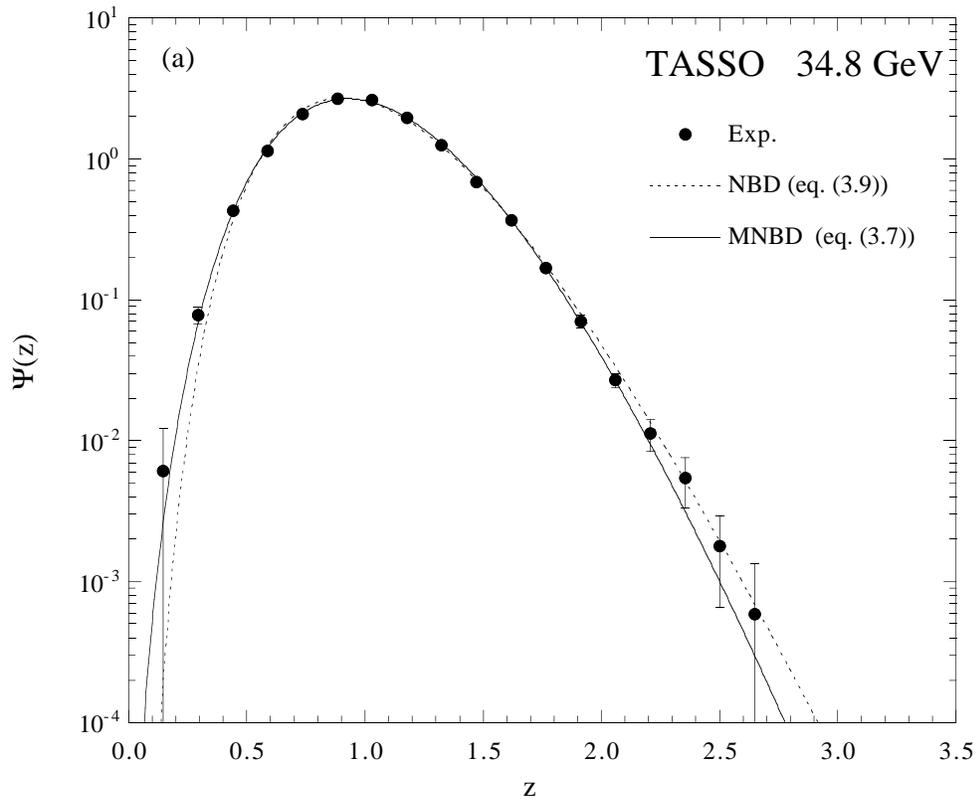
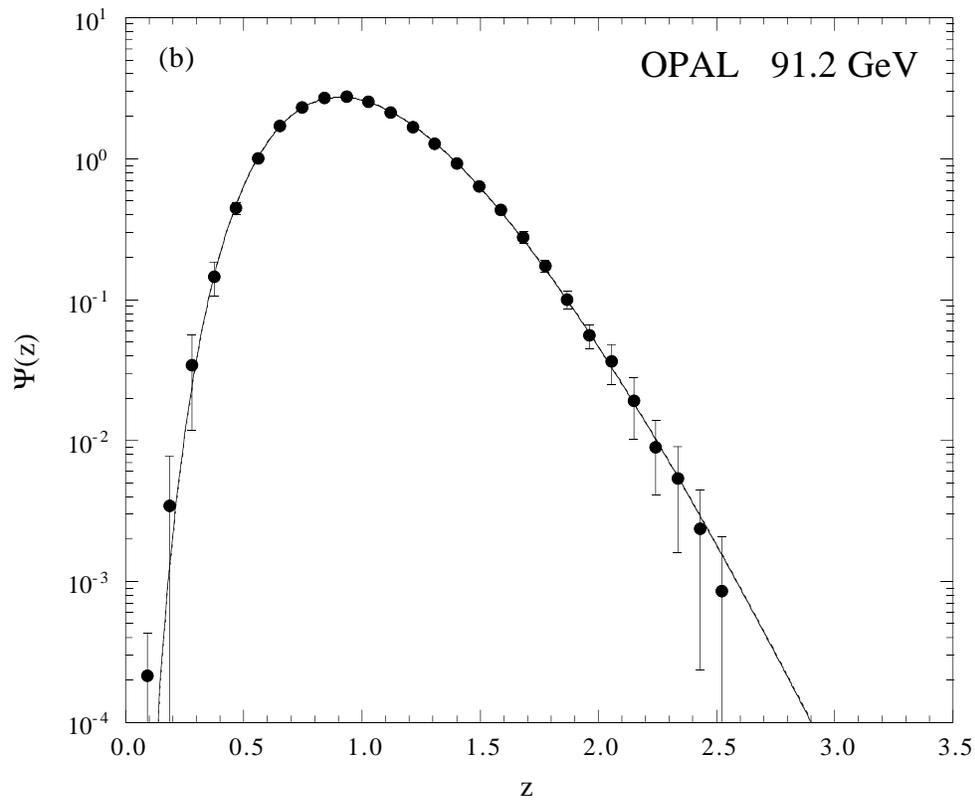
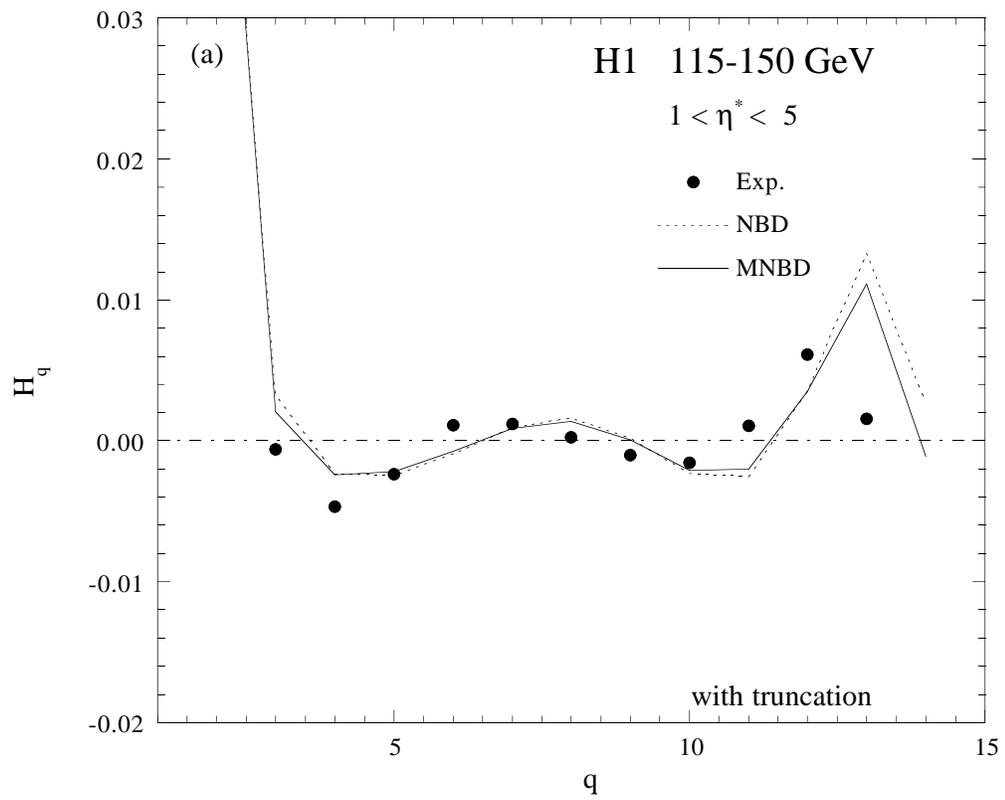
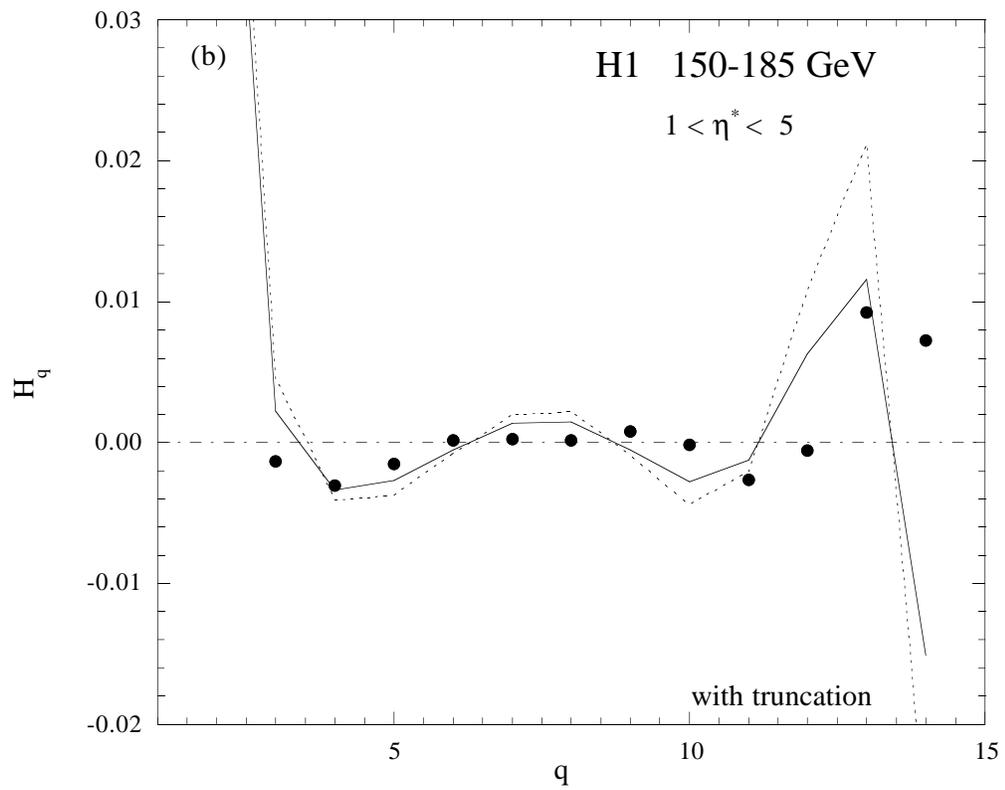
\end{document}